# Structural Stability, Electronic, Magnetic and Optical Properties of Rectangular Graphene and Boron-Nitride Quantum Dots: Effects of Size, Substitution and Electric Field

Sharma SRKC Yamijala [#, 1], Arkamita Bandyopadhyay [#, 2] and Swapan K Pati [2, 3, *]

*1. Chemistry and Physics of Materials Unit, Jawaharlal Nehru Centre for Advanced Scientific Research, Jakkur P.O., Bangalore – 560064, India.*

*2. New Chemistry Unit, Jawaharlal Nehru Centre for Advanced Scientific Research, Jakkur P.O., Bangalore – 560064, India.*

*3. Theoretical Sciences Unit, Jawaharlal Nehru Centre for Advanced Scientific Research, Jakkur P.O., Bangalore – 560064, India.*

*# These authors have contributed equally to this work*

*\* Corresponding author: pati@jncasr.ac.in*

## Abstract:

Using density functional theory calculations, we have examined the structural stability, electronic, magnetic and optical properties of rectangular shaped quantum dots (QDs) of graphene (G), Boron Nitride (BN) and their hybrids. Different hybrid QDs have been considered by substituting a GQD (BNQD) with BN-pairs (carbon atoms) at different positions. Several parameters like size, amount of substitution etc. have been varied for all these QDs (GQDs, BNQDs, hybrid-QDs) to monitor the corresponding changes in their properties. Among the considered parameters, we find that substitution can act as a powerful tool to attain interesting properties with these QDs, for example, broad range of absorption (~2000 nm) in the near infrared (NIR) region, spin-polarized HOMO-LUMO gaps without the application of any external-bias etc., which are highly required in the preparation of opto-electronic, electronic/spintronic devices etc. Explanations have been given in details by varying different factors, like, changing the position and amount of substitution, application of external electric-field etc., to ensure the reliability of our results.

## Introduction

Since the experimental discovery in 2004[1], graphene has become a subject of intense interest for scientists. Because of its unique properties,[2] graphene has been modified to use it in different devices.[3] But, pure graphene is a zero band-gap semiconductor or semi-metal,[2a, 2b] restricting its use into optoelectronic devices. This problem has partly been solved with the advent of quasi-one-dimensional graphene nanoribbons (the finite terminations of infinite graphene sheet) as they possess a band-gap.[4]

Band-gap of graphene nanoribbons (GNRs) have been proved to depend on their width.[4-5] Also, passivation and edge geometry (that is, zigzag or armchair or hybrid of zigzag and armchair) of a graphene nanoribbon changes its electronic, optical and magnetic properties.[4] According to the first principles calculations, both zigzag graphene nanoribbons (ZGNRs) and armchair graphene nanoribbons (AGNRs) are semiconducting.[5] Besides, ZGNRs have also been shown to change their electronic structure from semi-conducting to half-metallic (that is conducting for only one spin channel) upon the application of an external electric-field across the ribbon width.[6] Similar to ZGNRs, their inorganic and isoelectronic analogues, zigzag Boron-Nitride nanoribbons (ZBNNRs) have also attracted huge attention because of their intrinsic half-metallic nature[7] and also because of their several other interesting applications.[8]

Interestingly, several studies have shown that the electronic and magnetic properties of GNRs can be tuned either by substituting the GNRs' carbon atoms with B, N or by doping the GNRs with B/N atoms.[9] For example, doping the edge carbon atoms of a zigzag graphene nanoribbon with boron atoms, change the system into an intrinsic half-metal without application of any external electric field.[9b] Similarly, doping with B and N atoms in different ratios[9a] has shown to produce intrinsic half-metallicity in GNRs. Above findings clearly suggests that (a) reducing the dimensionality is one of the simple way to generate band-gap in graphene related materials and (b) combining graphene materials with BN would give interesting properties like half-metallicity to these hybrid materials.

By reducing one more dimension of GNRs, i.e. by generating zero-dimensional (0-D) graphene quantum dots (GQDs), one should expect a further opening in the band-gap because, unlike GNRs which are confined in two-dimensions, GQDs are confined in all the three dimensions. Indeed, recent reports have proved that there is a generation of the band-gap (more precisely, HOMO–LUMO gap) in GQDs.[10] Thus, GQDs can be considered as the bridge between small poly aromatic hydrocarbon (PAH) molecules and GNRs.

GQDs are found to possess unique electronic[10a, 10b], magnetic[10b] and optical[11] properties. Because of their tuneable energy gap, GQDs have been used in solar cells[12] and LEDs.[13] Additionally, GQDs are being emerged as the new carbon based graphitic fluorescent materials, which, depending upon their size and/or passivation can emit light of different wavelengths.[11b, 11c, 14] Along with their interesting optical properties, GQDs have also been used as bio-markers[15] because of their chemical inertness, biocompatibility and low-toxicity. *In vitro* studies have already been performed for imaging the cells using GQDs.[10d, 16] Furthermore, similar to GNRs, GQDs' properties are also shown to be dependent on their size and shape.[17] For example, smaller quantum dots have discrete energy levels, whereas, considerably larger nano-flakes have continuous energy bands[17b] and triangular nano-flakes have finite magnetic moments,[17a, 17d, 17e] etc. Similarly, properties of BNQDs have also been shown to vary with their shape and size.[17c]

Thus, there are studies, both experimentally and theoretically, on the effects of size, shape etc. of both GQDs and BNQDs. But, until now, there is no study which has investigated either the substitution or the doping of GQDs or BNQDs. Also, there is no study concentrating on the effects of the size, application of electric-field etc. on the electronic, magnetic and optical properties of GQD and BNQD hybrids. Moreover, hybrid BNC sheets have already been prepared experimentally with precise control over (a) the ratio of C:BN[18] and (b) domain shape of BN on C or vice-versa.[19] In addition, recent techniques like nanotomy[20] have shown to produce GQDs of desired geometries from graphite itself. Combining these ideas, we hope that one can prepare hybrid quantum dots of required shape, size and doping ratio, as required for better opto-electronic applications.

In the present study, we have performed spin-polarized first principles calculations on rectangular graphene (G), boron-nitride (BN) and hybrid quantum-dots (QDs). We considered our systems to be zigzag because it is known that producing ZGQDs are more favourable.[20-21] We have varied their width and the length to find out how/whether we can tune the HOMO–LUMO (H-L) gap and other properties of these ZGQDs by changing their size. Then, we have substituted these GQDs and BNQDs with BN and C, respectively, to find out the effects of substitution. As GNRs and BNNRs show several interesting electronic and magnetic properties including half-metallicity in the presence of electric-field,[6a, 7a] we have also studied the changes in the electronic and magnetic properties of the QDs in the presence of external electric-field. Unlike the previous studies, we have considered the application of electric-field along the diagonal direction, for the reasons which will be discussed later. We also have calculated the optical absorption spectra of different QDs, mainly, to find how the maximum absorption ($\lambda_{max}$) of GQD, BNQD and hybrid QDs differ from each other and also to know whether these QDs can be useful in any opto-electronic applications.

## Computational Details

Spin polarized first-principles calculations have been performed, to obtain all the electronic and magnetic properties of the systems, using the density functional theory (DFT) method as implemented in the SIESTA package[22]. Generalized gradient approximation (GGA) in the Perdew–Burke–Ernzerhof (PBE)[23] form has been considered for accounting the exchange-correlation function. Double ζ polarized (DZP) numerical atomic-orbital basis sets have been used for H, B, C, and N atoms. Norm-conserving pseudo-potentials in Kleinman-Bylander form[24] with 1, 3, 4 and 5 valence electrons for H, B, C, and N, respectively, have been considered. A reasonable mesh cut-off of 400 Ry for the grid integration has been used to represent the charge density and a vacuum of 20 Å has been maintained, around the quantum-dots, in all directions to avoid any spurious interactions. Systems are considered to be optimized if the magnitude of the forces acting on all atoms is less than 0.04 eV/Å. As the systems are zero-dimensional, all the calculations are performed only at the gamma ($\Gamma$) point of the Brillouin zone. Finally, optical properties have been calculated, for the optimized structures obtained from the SIESTA calculations, using the time dependent density functional theory (TDDFT) method as implemented in the Gaussian 09[25] package with hybrid B3LYP (Becke exchange with Lee, Yang and Parr correlation) functional[26] and with the basis-set 6-31+g(d).

## Results and Discussions

### A. Systems under consideration

In this work, we have considered rectangular quantum-dots (QDs) of graphene (G), boron-nitride (BN) and their hybrids with a zigzag edge along the direction of the length and an armchair edge along the width direction. Following the convention of the graphene nanoribbons,[4] we have assigned an ordered pair (n, m) to represent the length and width of the QDs. Here 'n' represents the number of atoms along the zigzag-edge (length) and 'm' represents the number of atoms along the armchair-edge (width), as shown in figure 1. In the present study, we have considered two values viz., 21 (~ 2.2 nm) and 33 (~ 4.2 nm) for 'n' and three values viz., 4 (~ 1 nm), 6 (~ 1.4 nm) and 8 (1.8 nm) for 'm'. Thus, for each type of QD (i.e. either a GQD or a BNQD or a hybrid) we have considered six different systems, each of which can be distinguished by the difference in their length × width.

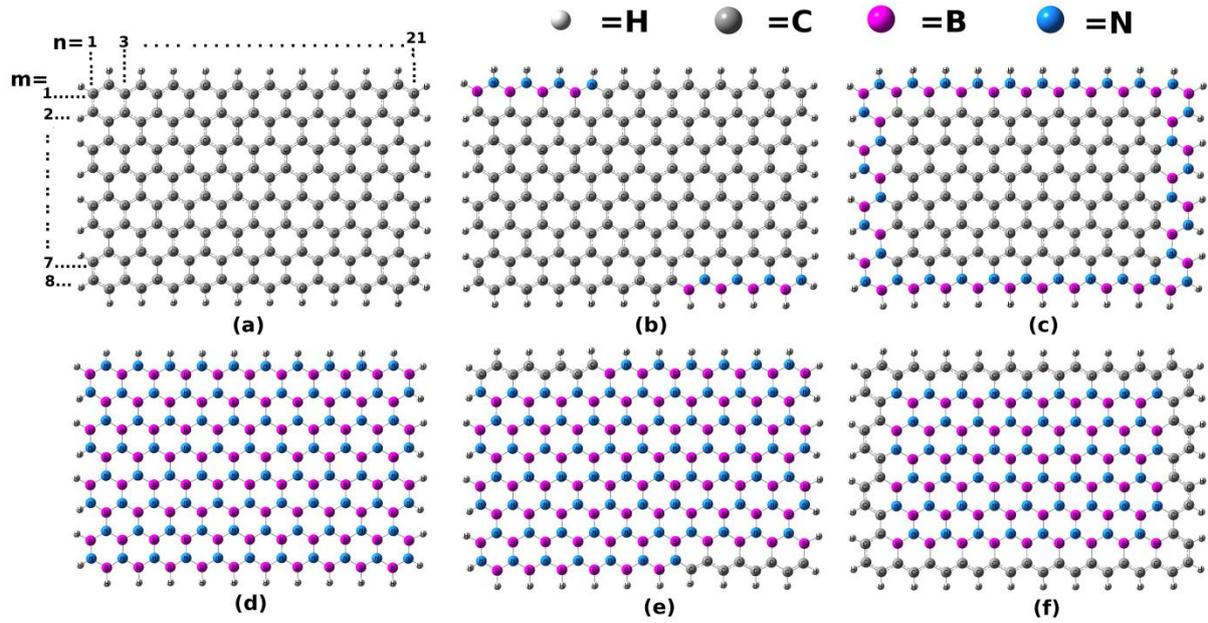

**Figure 1:** Structures of all the (21, 8) QDs. (a) GQD, (b) BN-partial-ed-GQD (c) BN-full-ed-GQD (d) BNQD (e) C-partial-ed-BNQD and (f) C-full-ed-BNQD.

As substitution on the pure QDs (i.e. GQDs and BNQDs) can be performed in several different manners (like changing the position, concentration etc. of the substituent atoms) and as it will be very difficult to study all types of substitutions, in the present study, we have mainly considered two ways of substitution viz., full-edge substitution (full-ed) and partial-edge substitution (partial-ed), as shown in figures 1b and 1c, respectively [also see SI]. If the GQDs or BNQDs have substitution, then the information of the substitution has been given first, followed by the name of the system. For example, a GQD (BNQD) whose edges are completely (partially) substituted with B and N atoms (C atoms) has been denoted as BN-full-ed-GQD (C-partial-ed-BNQD).

Coming to the passivation, we have passivated all our systems with H atoms because it has been proved that non-passivated edges are less stable than the passivated edges for the GNRs.[27,28] In order to avoid any confusion, we have also calculated the properties without passivation (just for one system–(21, 4) GQD) and we found that passivated QDs are more stable than the non-passivated QDs, in agreement with the previous studies.[27-28] Interestingly, from our spin-polarized calculations, we found that all the systems are stable with in the anti-ferromagnetic spin-configuration, following the Lieb's theorem.[29]

**Table 1:** Formation energies (eV) of the systems (21, 4), (21, 6) and (21, 8). Numbers inside the brackets (in italics) are the $(N_B + N_N) : (N_C)$ ratios.

| System Name | (21, 4) | (21, 6) | (21, 8) |
|---|---|---|---|
| GQD | -61.60 (*0 : 84*) | -70.62 (*0 : 126*) | -79.56 (*0 : 168*) |
| BN-partial-ed- GQD | -79.52 (*16 : 68*) | -88.48 (*16 : 110*) | -97.92 (*16 : 152*) |
| BN-full-ed- GQD | -126.56 (*50 : 34*) | -143.78 (*58 : 68*) | -163.81 (*66 : 102*) |
| C-full-ed-BNQD | -107.52 (*34 : 50*) | -170.64 (*68 : 58*) | -234.55 (*102 : 66*) |
| C-partial-ed-BNQD | -166.88 (*68 : 16*) | -244.44 (*110 : 16*) | -322.32 (*152 : 16*) |
| BNQD | -196.12 (*84 : 0*) | -274.92 (*126 : 0*) | -352.92 (*168 : 0*) |

**Table 2:** H-L gap (eV) of each system is given in both the spin-configurations (here after, we call one spin-configuration as spin-A and the other as spin-B).

| System Name | (21, 4) | | (21, 6) | | (21, 8) | | (33, 4) | | (33, 6) | | (33, 8) | |
|---|---|---|---|---|---|---|---|---|---|---|---|---|
| | spin-A | spin-B | spin-A | spin-B | spin-A | spin-B | spin-A | spin-B | spin-A | spin-B | spin-A | spin-B |
| GQD | 0.71 | 0.71 | 0.68 | 0.68 | 0.54 | 0.54 | 0.70 | 0.70 | 0.60 | 0.60 | 0.54 | 0.54 |
| BN-partial-ed- GQD | 0.81 | 0.16 | 0.04 | 0.54 | 0.04 | 0.43 | 0.05 | 0.51 | 0.32 | 0.04 | 0.01 | 0.19 |
| BN-full-ed- GQD | 0.71 | 0.71 | 0.41 | 0.11 | 0.09 | 0.34 | 0.62 | 0.62 | 0.32 | 0.06 | 0.22 | 0.03 |
| C-full-ed-BNQD | 0.28 | 0.28 | 0.44 | 0.44 | 0.48 | 0.48 | 0.05 | 0.23 | 0.16 | 0.16 | 0.19 | 0.19 |
| C-partial-ed-BNQD | 1.35 | 1.35 | 1.08 | 1.08 | 0.91 | 0.91 | 0.79 | 0.79 | 0.52 | 0.52 | 0.36 | 0.36 |
| BNQD | 4.32 | 4.32 | 4.16 | 4.16 | 4.03 | 4.03 | 4.32 | 4.32 | 4.16 | 4.16 | 4.06 | 4.06 |

## B. Stability

Stability of a system has been assessed by its formation energy ($E_{Form}$). $E_{Form}$ has been defined as:

$$E_{Form} = (E_{tot} - N_C E_C - N_H E_H - N_B E_B - N_N E_N),$$

where, $E_{tot}$, $E_H$, $E_B$, $E_C$ and $E_N$ are the total energy of the system, energy of a H atom (calculated from a hydrogen molecule), energy of a B atom (calculated from α-boron), energy of a C atom (calculated from a 8 × 8 graphene super-cell) and energy of a N atom (calculated from a nitrogen molecule), respectively and $N_H$, $N_B$, $N_C$ and $N_N$ are the number of H, B, C and N atoms present in the system, respectively. $E_{Form}$ of the systems (21, 4), (21, 6) and (21, 8) are listed in table 1. Formation energies of all the systems are negative (see table 1), which indicates that all these systems are thermodynamically feasible. From the table we find that, for a particular QD, with an increase in the size its $E_{Form}$ becomes more negative, i.e. the QD becomes more stable. This is expected because, with an increase in the system size the ratio of the number of atoms at the edge (more reactive/less stable) to the number of atoms at the bulk (less reactive/more stable) will decrease, and hence, the system will get more stabilized. In fact, from the $E_{Form}$ values it appears that, BNQDs are more stable than GQDs and the $E_{Form}$ values of the QDs are in the order of: GQD > BN-partial-ed-GQD > BN-full-ed-GQD > C-full-ed-BNQD > C-partial-ed-BNQD > BNQD for all the sizes (also see SI).

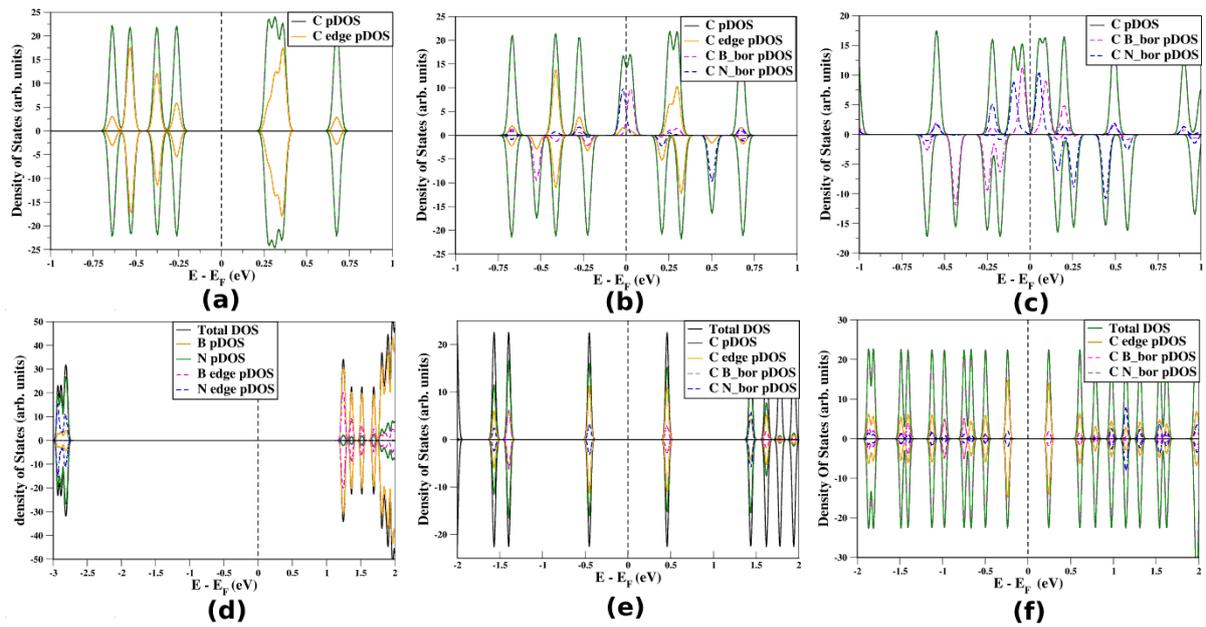

**Figure 2:** pDOS plots of all (21, 8) QDs. (a) GQD, (b) BN-partial-ed-GQD (c) BN-full-ed-GQD (d) BNQD (e) C-partial-ed-BNQD and (f) C-full-ed-BNQD.

**C. Electronic and magnetic properties**

In this section, we will present the electronic (density of states (DOS), projected DOS (pDOS), H–L gap, charge-transfer and wave-functions) and magnetic properties of the rectangular QDs. We will first explain the changes in these properties with the size, then with the substitution, and finally with the application of an external electric-field.

*Effect of Size:*

Similar with the case of GNRs,[30] GQDs also show a finite H-L gap because of the quantum confinement effect (here, in both the directions). As shown in the table 2, for all the QDs there is a decrement in the H-L gap with an increment in the length and/or width of the QDs. For example, (21, 4) GQD has a H-L gap of 0.71 eV for both the spin-channels, but, once the system size has been increased to (21, 6) and (21, 8), there is a decrement in the H-L gap (from 0.71 eV) to 0.68 and to 0.54 eV, respectively. Lesser H-L gap for greater sizes is due to the (a) decrement in the energy-level spacing, (b) increment in the delocalization of the π electrons (due to the increased $sp^2$ bonded region) in the larger QDs. The H-L gap of 0.54 eV for the (21, 8) or (33, 8) is comparable with the calculated band-gap of 0.51 eV for the 8-ZGNR obtained within the LDA-approximation.[5] Comparable results have also been obtained previously by Philip Shemella *et al.,*[31] and Oded Hod *et al.*[10a] Similar to GQDs, all the other QDs (considered in this study) have shown a decrement in the H-L gap with an increment in their size (also see SI).

Figure 2a shows the density of states (DOS) and projected-DOS (pDOS) of (21, 8) GQD. Clearly, DOS (green color) near the Fermi-level has a major contribution from the zigzag-edge atoms (orange colour), and also, both the HOMO and LUMO of (21, 8) GQD shows that the wave-functions are localized at the zigzag-edges (see figure S2). One notable point is HOMO and LUMO are localized at two different edges for each spin, and also, the edge where the wave-function is localized for the spin-A's HOMO is the same as the spin-B's LUMO and vice-a-versa. Similar localization behaviour of wave-function has been previously observed in ZGNRs studied by Zheng *et al*.[32] When we changed the widths and lengths of GQDs, we find similar localization behaviour except that the degree of localization of the wave-function at a particular edge increases with a decrement in the width of the GQD. This result is in contradiction to Shi *et al.*[33], although their findings are mainly for non-passivated GQDs,[34] but compares fairly well with the Hod *et al.*[10a] As the DOS near the Fermi-level is mainly contributed by the zigzag edges and as substitution at the zigzag edges (with a similar kind of localization behaviour of wave-function[32] which we find for GQDs) has proven to be useful to attain several interesting properties like half-metallicity in ZGNRs [9a, 9c, 35], we have substituted these edges with the isoelectronic BN-pairs.

**Table 3:** Mülliken population analysis of (21, 4)-GQDs (both pristine and substituted). Amount of the charge-transfer from (to) the edge or border carbon atoms to (from) the substituted nitrogen or boron atoms is given in the last row, for each spin, separately.

| GQD | | | | BN-zigzag-ed-GQD | | | | BN-partial-ed-GQD | | | | | | | |
|---|---|---|---|---|---|---|---|---|---|---|---|---|---|---|---|
| Avg. edge | | Avg. 2nd zigzag line | | B-border | | N-border | | B-edge | | N-edge | | B-border | | N-border | |
| spin-A | spin-B | spin-A | spin-B | spin-A | spin-B | spin-A | spin-B | spin-A | spin-B | spin-A | spin-B | spin-A | spin-B | spin-A | spin-B |
| 19.02 | 19.02 | 21.48 | 21.49 | 20.81 | 20.81 | 22.26 | 22.26 | 11.99 | 10.82 | 10.85 | 12.01 | 7.21 | 7.81 | 8.38 | 7.81 |
| Charge transfer | | | | -0.67 | -0.67 | +0.78 | +0.78 | +0.58 | -0.59 | -0.56 | +0.6 | -0.61 | +0.01 | +0.57 | -0.002 |

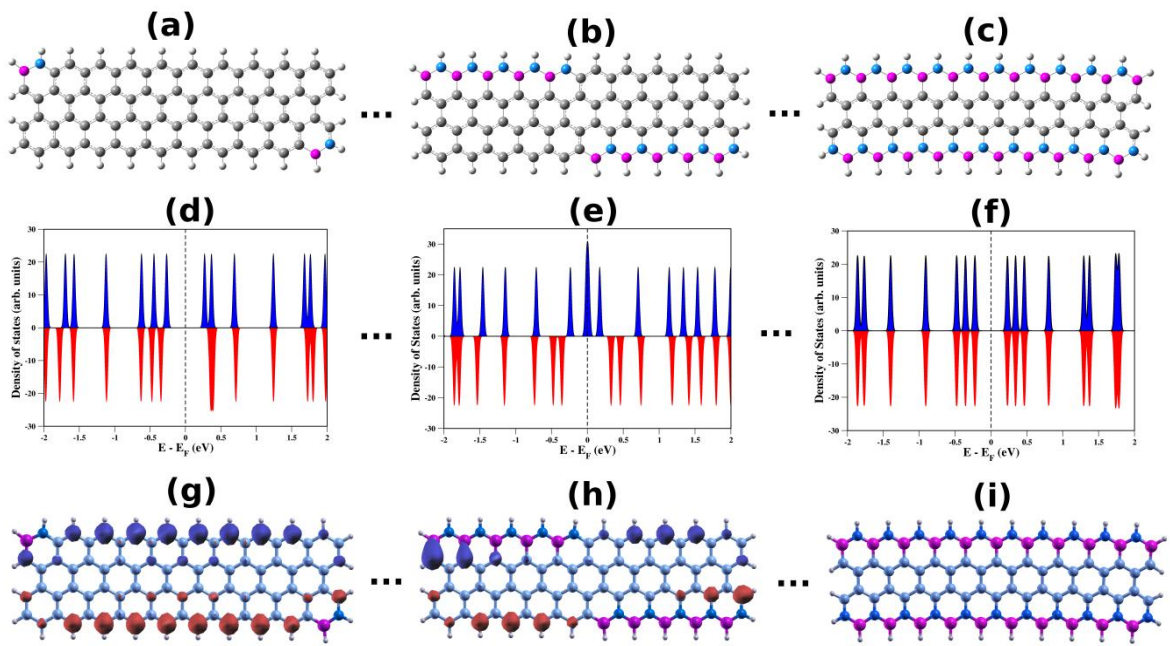

**Figure 3:** Variation in the spin-dependency of the H-L gap with a variation in the amount of substitution at the zigzag edges of (21, 4)-GQDs. (a-c) shows structures with different amount of substitution; (d-f) shows their corresponding DOS and (g-i) are their corresponding spin-distributions. Structures, DOS plots and spin-distributions for all the other substitution levels considered in this study were given in figures S6, S7 and S8, respectively.

### *Effect of Substitution:*

We find that, when we substitute a GQD's edge completely with BN, the H–L gap of the system decreases. Interestingly, when the width of the QD is greater than 4, we find a spin-dependent H–L gap (see table 2 and figures S3d-S3f). The spin-dependency of the H–L gap is because of the intrinsic electric-field present in the system. In fact, due to the presence of B and N-atoms at two different zigzag-edges and as B is a Lewis acid, there would be a charge-transfer which generates the potential gradient across the ribbon.[35] The reason for the spin-independency of the H–L gap of the BN-full-ed-GQDs with width 4 is solely because of their smaller width. With a decrease in the width of the QD, the strength of the field which is required to lift the degeneracy between the two configurations of the spins will also increase. Similar explanation regarding the width dependency of the critical electric-field to break the degeneracy of the two spin configurations for the case of GNRs has been discussed by Son *et al*.[6a] Additionally, partial BN substitution on GQDs also produces a spin-dependent H–L gap (see table 2 and figures S3a-S3c), similar to complete edge substitution, but, the *spin-dependency is there for all the widths*. This result is quite interesting and by bringing the analogy from the previous studies on the ZGNRs,[6a, 32] ZBNNRs[7] and their hybrids[9c] (also see SI) we find that there are several possible ways to explain the spin-polarization in BN-partial-ed-GQDs, namely, presence of localized edge-states due to edge carbon-atoms, presence of an intrinsic electric-field along the diagonal direction, presence of border carbon atoms which can strongly contribute to the states near the Fermi-level etc.

Among the above, from our calculations, we find that the main reason for the spin-dependent H-L gap is due to the presence of both border and edge carbon atoms in BN-partial-ed-GQDs (see SI to know how we discarded the other reasons). To prove it, we have gradually increased the amount of substitution at both the zigzag edges of (21, 4)-GQDs, as shown in figure 3a. From this, we find that the H-L gap of (21, 4)–GQD is spin-dependent for all substitutions except for the complete edge substitution (see figure 3b). To further prove that this spin-dependency in H-L gap is occurring solely because of the edge and border carbon atoms, we have also plotted the spin-distributions in these systems and are shown in figure 3c. These spin-distributions show us that there is a high spin-non-degeneracy at the border and the edge-carbon atoms. Also, to show the effect of width dependency, we have plotted the pDOS of BN-partial-ed-GQDs at different widths (see figures S3a-S3c). From these figures, it is clear that the DOS near the Fermi-level always has a major contribution from the border and edge carbon atoms.

To further understand the origin of this spin-dependency, we have performed the Mülliken population analysis and the results for (21, 4)-GQDs (for both pristine and substituted) are tabulated in table 3. Clearly, the average amounts of charge (total charge at each edge, for each spin, divided by 2) present at the edge and at the 2nd zigzag line of a GQD are same for both the spins, and hence, the H-L gap is spin-degenerate. [In a GQD, the charge present at each edge is not same for both the spins but the amount of the charge in up-spin at one-edge will be equal to the amount of charge in down-spin for the other-edge, very similar to ZGNRs. Same rule applies for $2^{nd}$ zigzag line also]. In BN-full-ed-GQD, although the amount of charge transfer to the border carbon atoms at B-border is different to that of N-border, there is no difference in amount of the charge transferred through each spin, and hence, again the H-L gap is spin-degenerate. But, when it comes to the BN-partial-ed-GQD, there is an apparent difference also in the amount of the charge-transferred through each spin at each border. In other words, "In a BN-partial-ed-GQD, there is an inherent difference in the amount of charge transferred through spin-A to that of spin-B, whenever there is a charge transfer from a Boron/Nitrogen atom to a border-carbon atom". Similar charge transfer behaviour has been observed for other sizes of BN-partial-ed-GQDs and the values are given in table S1. Thus, the origin of the spin-dependent H-L gap lies in the spin-dependent charge-transfer to the border carbon atoms and is manifested in the DOS and pDOS plots. Finally, it is important to mention that in all the above substitution studies, we have placed boron atoms at one edge and nitrogen atoms at the other, which leads to two different kinds of edge carbon atoms, and hence, also two different kinds of border carbon atoms. This is another important reason for the observation of the spin-dependent H-L gap in BN-parital-ed-GQDs even for small widths (see SI for further details).

In BNQDs, substitution (both full-edge and partial-edge) decreases their H-L gaps. Surprisingly, H-L gap of C-full-ed-BNQDs is even below the H-L gap of GQDs, for all widths and lengths. Here also, it occurs solely because of the substituted edge carbon atoms, as shown in the pDOS plots (see figures 2e and 2f).

From all the above we can conclude that, one can achieve a spin-polarized H-L gap in a rectangular GQD if any of the following conditions is satisfied:

(i) Complete substitution of the zigzag edges of GQD with BN pairs, for widths greater than 1 nm.
(ii) Partial zigzag edge substitution (i.e. irrespective of number and position) of a GQD with BN pairs, in such a manner that, nitrogen atoms are at one zigzag edge and boron atoms are at the other.

Above conclusions have been verified up to a size of ~ $4.2 \times 2.2$ nm$^2$ and we expect them to work for higher sizes (than what we have considered here) also.

*Effect of external electric field on GQDs:*

In case of BN-partial-ed-GQDs, it is not the intrinsic electric-field which gives rise of spin-polarization. To prove this, first we have to show that GQDs are spin-polarized (see SI for further details) when an external electric-field acts along the diagonal direction (because charge-transfer effects are present along the diagonal direction in BN-partial-ed-GQDs) of the GQDs. Results of these calculations for (21, 4) GQD are given here. To ensure the stability of the system under the applied electric-field, first we have calculated the minimum force (when applied between the diagonal edges of GQD) at which the GQD structure starts destroying. The minimum force, $F_{min}$, is calculated as,

$$F_{min} = (\text{C-C bond-energy of GQD}) / (\text{Distance between the diagonal edges}).$$

For (21, 4)-GQD the distance between the diagonal edges is 27.49 Å and we took the C-C bond-energy to be 4.9 eV.[36] Substituting these values gives us $F_{min} \approx 0.18$ eV/Å. Thus the minimum strength of the electric-field at which the destruction of the GQD structure starts is $E_{max} \approx 0.18$ V/Å. Considering this $E_{min}$ value, we have applied an external electric field, only in the range of 0.01 – 0.05 V/Å, i.e. well below the $E_{min}$ value, across the diagonal of the rectangular (21, 4) GQD. We find that, in the range of applied electric-field, the H-L gap of (21, 4)-GQD is always spin-dependent as shown in the inset of figure 4(i). The spin-dependency in the H-L gap of GQDs can simply be explained from the fact that the applied electric field shifts the energy levels of the opposite spins in different directions, thus, breaking the localized edge-state spin-symmetry. A similar phenomenon was also observed when an external electric field was applied across the zigzag edges (though not along the diagonal direction) of a ZGQD.[10a] To further understand the reason for such spin-polarized H-L gap, we have presented the individual shifts of the HOMOs and LUMOs for both the spins in figure 4(i). From the figure it is clear that, energy of both the HOMOs (LUMOs) is increasing (decreasing) with an increase in the external electric-filed [as HOMO (LUMO) corresponds to electron (hole) occupied state]. These increments / decrements in the MO energies are consistent with the linear stark-effect, where the shift in the energy of a MO is given by $\Delta E = -\vec{P}\cdot\vec{E}$, where $\vec{P}$ is the polarization and $\vec{E}$ is the applied electric-field.[37] From the equation, it is clear that the shift in the MO energy is proportional to the polarization and the shift will be positive or destabilizing (negative or stabilizing) if the direction of the polarization is opposite (parallel) to the applied electric-field. Thus, the change in the energy of an MO due to the external electric-field can be understood by knowing the direction of polarization of that MO.

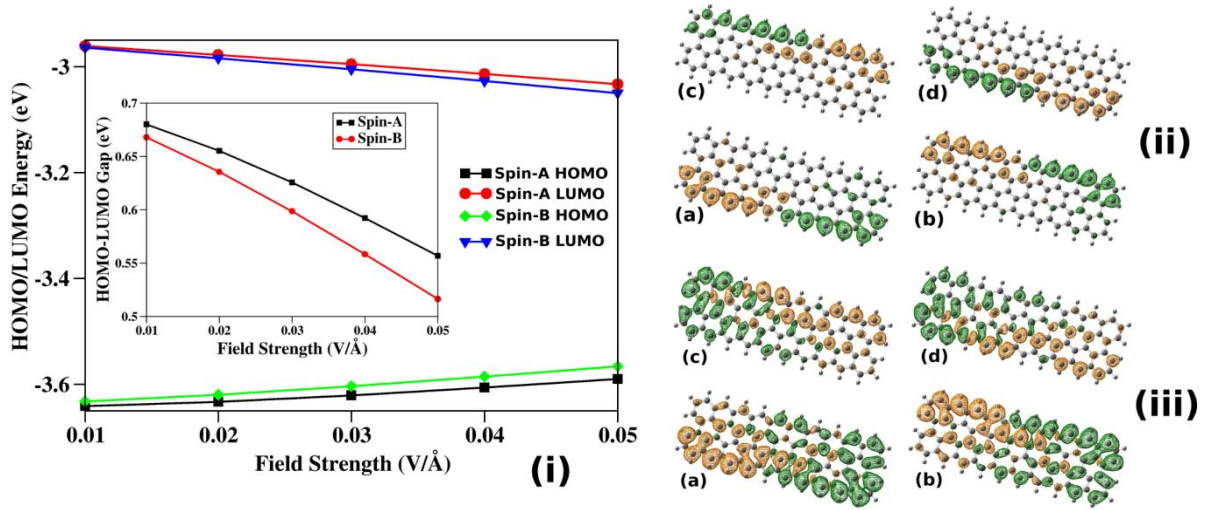

**Figure 4: (i)** Absolute energies of MOs of (21, 4)-GQD under the influence of an external electric-field. Inset shows the corresponding variation in the HOMO-LUMO gap. **(ii, iii)** Electron density difference maps (EDDMs) of (21, 4)-GQD. The two electric-fields considered for calculating EDDMs are 0.01 and 0.02 eV/Å for (ii) and 0.04 and 0.05 eV/Å for (iii) [see the text for details]. (a), (b), (c) and (d) in each sub-figure corresponds to spin-A-HOMO, spin-B-HOMO, spin-A-LUMO and spin-B-LUMO, respectively. For all the plots an isovalue of 0.002 e/(Å)$^3$ is considered. The direction of polarization is from the green region to orange region.

To get the direction of polarization of MOs, we have generated the cube files of each MO and took their squares to get the corresponding electron-densities of each MO. Then, we took the difference between the electron-densities of each MO at two different electric-fields. These electron-density difference maps (EDDMs) are given in figure 4(ii, iii), where the green area shows the loss in the electron density and the orange area shows the gain in the electron density, and hence, the direction of polarization is always from green to orange. Now, from figure 4(ii, iii), the reason for the shift in the MO energies is very obvious – a positive (negative) shift if the direction of polarization is opposite (parallel) to the applied electric-field direction. Thus, the shifts in the MO energies are consistent with the linear Stark-effect in the applied range of external electric-field. So, we conclude that, (i) similar to the ZGNRs,[6a] H-L gap of GQDs can be tuned with an external electric-field (ii) these changes in the H-L gap are consistent with the linear Stark-effect, (iii) such changes can be understood from the EDDMs plots, and importantly, (iv) although the H-L gap of GQDs can be spin-polarized under external electric-field, the spin-dependent H-L gap of BN-partial-ed-GQDs is not due to the internal electric-field, for the reasons discussed earlier.

### D. Optical properties:

Using TDDFT, we have calculated the absorption properties of (21, 4)-QDs. Clearly, all QDs absorb in the UV-Visible region (figure 5), although, the wavelength of maximum absorbance ($\lambda_{max}$) is in this region only for the BNQDs (both pure and substituted (figure 6b)). Among the BNQDs, pure BNQD has its $\lambda_{max}$ in the UV region, but, its oscillatory-strength (0.07) is very low (also see table S3) compared to any other QDs. $\lambda_{max}$ values of both C-partial-ed-BNQD (~ equal oscillator strength peaks [0.81 & 0.79] at 427, 440 nm) and C-full-ed-BNQD (~ 618 nm) are in the visible region, with good oscillatory strength values of ~0.8 and ~2.8, respectively. Electronic transitions whose contributions are large for these peaks are given in table S3 and from our EDDMs plots (figure S9) we infer that these high oscillatory strength peaks corresponds to charge-transfer excitations in both C-partial-ed-BNQD and C-full-ed-BNQD. In C-full-ed-BNQD, the charge transfer is different at different edges – at N-edge, charge has transferred from edge carbon atoms to border carbon atoms, whereas, at B-edge it has transferred from border to edge carbon atoms. In C-parital-ed-BNQD, for the 427 peak, charge has transferred from border carbon atoms to edge carbon atoms and for 440 peak, we find the opposite trend. Thus, all these high oscillatory strength peaks corresponds $\pi$ - $\pi^*$ transitions and are of charge-transfer type (mainly involving border/edge C-atoms).

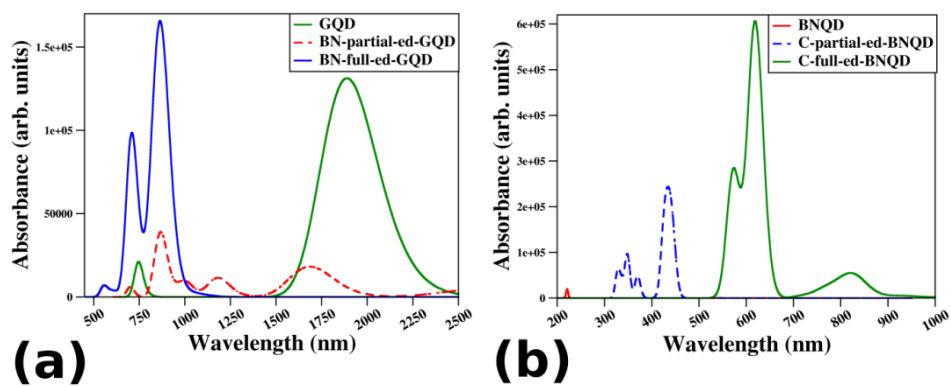

**Figure 5:** Optical absorption profiles of (a) GQD, BN-partial-ed-GQD and BN-full-ed-GQD; (b) BNQD, C-partial-ed-BNQD and C-full-ed-BNQD calculated using TDDFT.

Among GQDs, we find that pure GQD has its $\lambda_{max}$ in the near-IR (NIR) region (1878.38 nm), but its absorption profile spans over ~ 1000 nm. Absorbing in IR region together with this wide-range suggests that it could be a potential candidate for the preparation of solar cells mainly because: (i) half of the Sun's energy is emitted in the IR-region[38] and (ii) most of the present day solar-cells are not efficiently utilizing the red and IR regions of solar spectrum.[39] Though not as intense as GQD, BN-partial-ed-GQD also has a wide range absorption profile. In fact, the absorption profile of BN-partial-ed-GQD spans more than twice to that of the GQD absorption profile (i.e. > 2000 nm), and again, most of it is in the NIR-region. Unlike GQDs and BN-partial-ed-GQDs, BN-full-ed-GQDs have a strong absorption also in the visible-region of the spectrum (apart from the strongest peak in the NIR-regions). From the EDDMs plots (see figure S9) we find that all the peaks arise due to $\pi$ - $\pi^*$ transitions and the most-intense peaks of all the QDs are because of the charge-transfer. Thus, based on our TDDFT calculations, we expect that GQDs (both pure and substituted) can act as potential candidates for the preparation of active components of solar cells due to their wide range of absorption in the NIR-region.

## Conclusions:

In conclusion, we have examined the structural stability, electronic, magnetic and optical properties of rectangular GQDs, BNQDs and their hybrids. Also, changes in the properties of the quantum dots due to effects of size variation as well as substitution have been studied. All the systems were found to be thermodynamically stable and their stability was found to increase with an increase in the $(N_B + N_N)$: $(N_C)$ ratio. Increment in the sizes of the QDs lead to the decrement of the H-L gap in all the systems except for C-full-ed-BNQDs (where, the increased delocalization of the $\pi$-electrons on the C atoms made this system's HOMO more stable leaving its LUMO energy unchanged). Substituting the pure QDs gave several interesting properties. Above a critical width, BN-full-ed-GQDs were found to possess spin-polarized H-L gap due to the presence of an intrinsic electric-field in the system which can polarize the orbitals. Unlike BN-full-ed-GQDs, BN-partial-ed-GQDs have shown spin-polarized H-L gaps irrespective of the width of the system. We have clearly shown that electric-field is not the reason for the spin-polarized H-L gap in BN-partial-ed-GQDs, although it can create spin-polarized H-L gap in GQDs when applied along the diagonal direction of the QDs. After performing several studies by changing the position and the amount of substitution in GQDs, we found that the reason behind the spin polarized H-L gap in BN-partial-ed-GQDs is the spin polarized charge transfer between the border carbon atoms by boron/nitrogen atoms. As this behaviour is not observed when only one type of edge or border atom is there in the system, we have concluded that to attain spin-polarized H-L gap in a system, we must have different types of edge and border C atoms. Next, we have shown that an external electric field can induce spin-polarized H-L gap in GQDs and we explained the energy shifts of the MOs under external electric-field by plotting the EDDMs proving that our results are consistent with linear Stark-effect. Interestingly, we have studied the optical absorption properties of all the (21, 4)-QDs and we find that they can absorb light in a broad energy window ranging from IR to UV. We have also seen that some of these QDs have wide range absorption profiles in NIR region–the property which is required for a material to absorb more solar-energy– and this property of these QDs made us to conjecture that they can be selected as possible candidates for designing efficient solar-cells.

# Supporting Information

**Related to "A. Systems under consideration":**

We would like to emphasize that other than the two systems which we have studied here one could have considered any of the other possible substituted systems to understand the behaviour of hybrid QDs. The only point which should be noted here is, in the case of partial-edge substitution, we have intentionally substituted the zigzag edge (rather than the armchair-edge) because it has been proven that zigzag-graphene-nanoribbons can show interesting magnetic properties like half-metallicity,[6a, 9a, 9b] if the spin-degeneracy at the zigzag-edge can be broken [for example, using methods like substitution,[9c] doping,[9b] external electric-field[6a] or their combinations [9a]].

**Related to "B. Stability":**

The order of: GQD > BN-partial-ed-GQD > BN-full-ed-GQD > C-full-ed-BNQD > C-partial-ed-BNQD > BNQD is found for all the systems except for the (21, 4), where, we find the order: "C-full-ed-BNQD > BN-full-ed-GQD" instead of the expected order: "BN-full-ed-GQD > C-full-ed-BNQD". This is not an anomalous behaviour and can be explained as follows: As BNQDs are more stable than GQDs, it is obvious that, the stability of a system which has both BN and C (i.e. either full-ed or partial-ed systems) will be more if it has more number of B and N atoms than the C atoms. In table 1, we have shown the ratio between the total number of boron and nitrogen atoms ($N_B + N_N$) and the number of carbon atoms ($N_C$) for each system. From the table 1 it is clear that, among the C-full-ed-BNQD systems the ratio ($N_B + N_N$): ($N_C$) is less than one (i.e. $N_B + N_N < N_C$) only for the case of (21, 4) system, and hence, this system is found to be less stable than BN-full-ed-GQD.

**Related to "C. Electronic and magnetic properties":**

All the systems have a zero total-spin-polarization, as expected, because of their anti-ferromagnetic nature, but, some of the systems have a finite spin-polarization near the Fermi-level (for example, see figure 2b), and the reasons are the reasons are explained in the main article.

***Related to "Effect of Size":***

All the other QDs (considered in this study) have shown a decrement in the H-L gap with an increment in their size, *except for the C-full-ed-BNQD*. This is because as the width of the ribbon increases, in C-full-ed-BNQD, the number of carbon atoms present at the edge increases, which in-turn increases the delocalization of electrons (because B, N atoms are replaced with carbon-atoms). This delocalization brings in stability mainly to the HOMO (leaving the LUMO almost unchanged, see Figure S1), and hence, the increment in the H–L gap with an increment in the width of this QD.

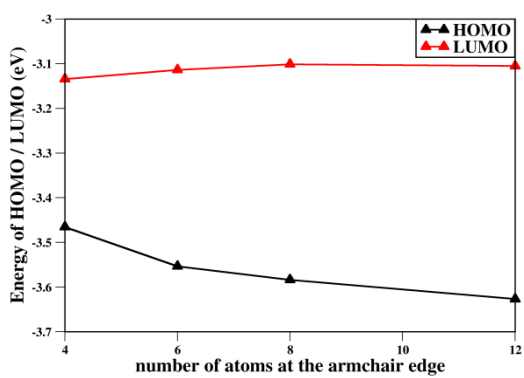

**Figure S1:** HOMO and LUMO energies of C-full-ed-BNQDs with different widths.

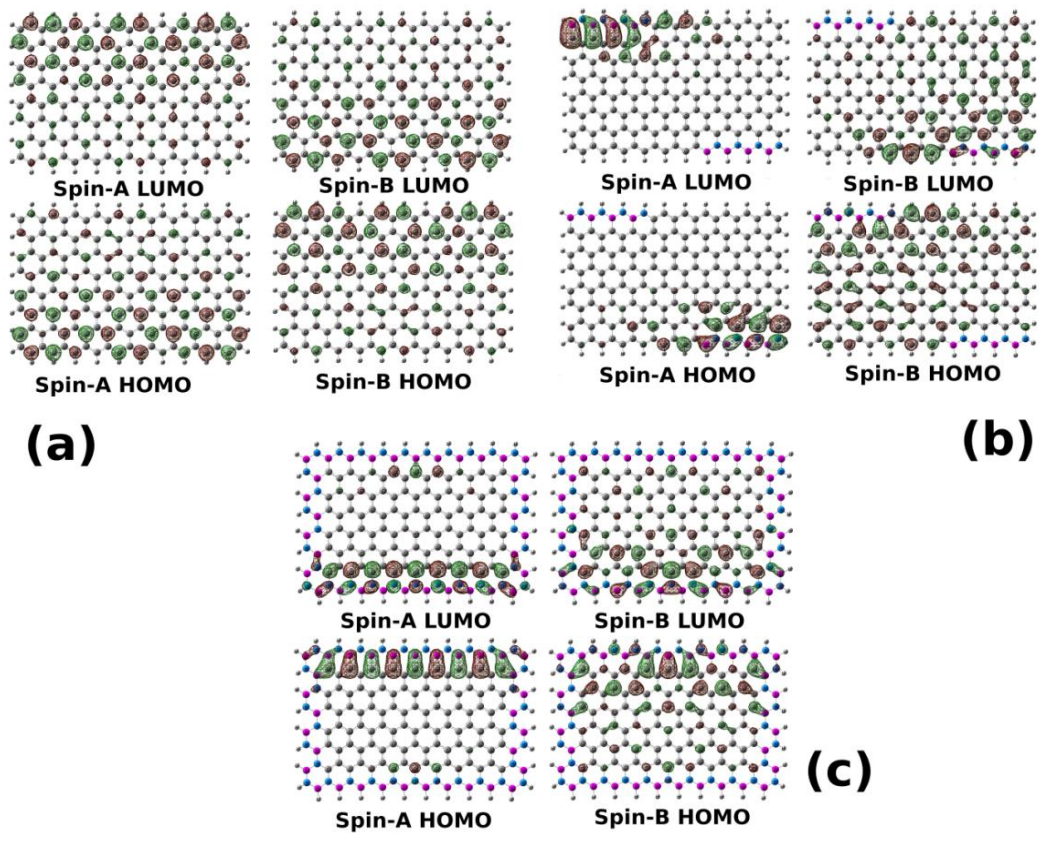

**Figure S2:** HOMO-LUMO plots of (21, 8) QDs. (a) GQD, (b) BN-partial-ed-GQD and (c) BN-full-ed-GQD

*Related to "Effect of Substitution":*

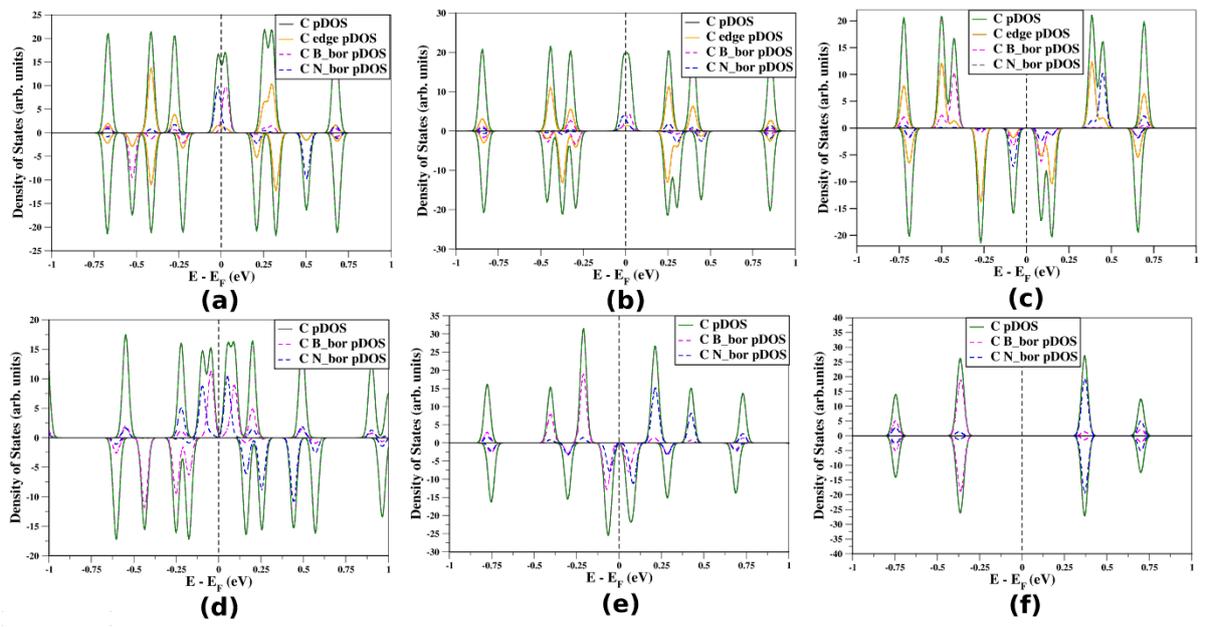

**Figure S3:** pDOS plots of (a) (21, 8) BN-partial-ed-GQD, (b) (21, 6) BN-partial-ed-GQD, (c) (21, 4) BN-partial-ed-GQD, (d) (21, 8) BN-full-ed-GQD, (e) (21, 6) BN-full-ed-GQD, (f) (21, 4) BN-full-ed-GQD.

Spin-polarized H-L gap in BN-Partial-ed-GQDS: This result is quite interesting, because, unlike the BN-full-ed-GQDs, the reason for this spin-polarization is not just due to the intrinsic electric-filed (as shown in the main article).

Based on the previous studies on the ZGNRs,[6a, 32] we noted that ZGNRs can be tuned to half-metals (i.e., the spin-degeneracy is lifted) by a shift in the energy of the two zigzag-edge states differently, which can be done using electric-field, doping, substitution etc. A similar methodology has also been applied to achieve half-metallicity in ZBNNRs, but, only when the zigzag edges are bare.[7] In contrast to both ZGNRs and ZBNNRs, half-metallicity in hybrid nanoribbons is not just because of the edge-atoms but because of the combined effect of both edge and border atoms (in particular, there is a larger contribution of border atoms near the Fermi-level).[9c] As BN-partial-ed-GQDs have all the above possibilities, we thought that the reason for their spin-dependent H-L gap could be due to the combined effect of some/all of them or because of any one of them.

*Proving that "Intrinsic electric-field is not the cause for spin-polarized H-L gap in BN-partial-ed-GQDs:*

As each edge of these systems possesses both the carbon atoms and B/N atoms, we have first checked for the presence of any *intrinsic electric-field* in these systems. Unlike the BN-full-ed-GQDs, here the intrinsic electric-field (if exists) can act only along the diagonal of the rectangular QDs. So, we have applied an electric-field (from 0.01 to 0.05 V/Å) across the diagonal of the GQDs and we find that H-L gap of all the systems are indeed spin-dependent (see "Effect of external electric field on GQDs" section of the main article for further details). This shows that, if an electric-field is applied along the diagonal direction, then the spin-degeneracy in the H-L gap of a GQD can possibly be lifted.

But, if the electric-field (here, intrinsic) is the main reason for spin-non-degeneracy, then (according to the previous arguments on the intrinsic electric-field[35]) one should observe spin-non-degeneracy only above a particular diagonal distance (For GNRs, it is the distance between two zigzag edges.[35]) and below which the H-L gap should be spin-degenerate. When we performed the calculations with a decreased diagonal distance (by decreasing width and/or length of the GQD and also by increasing the substitution), we couldn't find any spin-degeneracy, as shown in figure S4(i). Thus, although an electric-field if acted across the diagonal direction of GQD can cause a spin-dependent H-L gap, it is not the reason for the spin-dependent H-L gap in (21, 4)-BN-partial-ed-GQD. To support the above conclusion we have applied an external electric-field on (15, 4) and (20, 3) GQDs, across their diagonal direction, and we find that their H-L gaps are spin-independent for lower electric field and spin-dependent for higher electric field (see the discussions below figures S5 and S7).

From figure S5, it is clear that the strength of the electric-field required to polarize the H-L gap of the (21, 4) system (i.e. to make it spin-dependent) is less when compared to the strength of the field required to polarize (15, 4) system. At 0.01 V/Å, both the system's H-L gap is spin-degenerate. At 0.1 V/Å, H-L gap is spin-dependent only for (21, 4) system. And at 0.15 V/Å, both the system's H-L gap is spin-dependent. This clearly proves that, lesser the diagonal distance of the GQD higher the external electric-field strength required in achieving a spin-dependent H-L gap in a GQD and there is a critical field above which H-L gap of a GQD will be spin-dependent. Now, referring back to the figure S4, it can be noted that, though we have decreased the width of the system from (21, 4) to (15, 4), the spin-dependency of the H-L gap hasn't changed. As we couldn't find any dependency of the "spin-dependent H-L gap" on the size of the system in (21, 4)-BN-partial-ed-GQD despite of such dependency in (21, 4)-GQD under external electric-field, we concluded that intrinsic electric-field acting along the diagonal direction can't be the cause for the spin-dependent H-L gap of (21, 4)-BN-partial-ed-GQD (also see the discussion below figure S7).

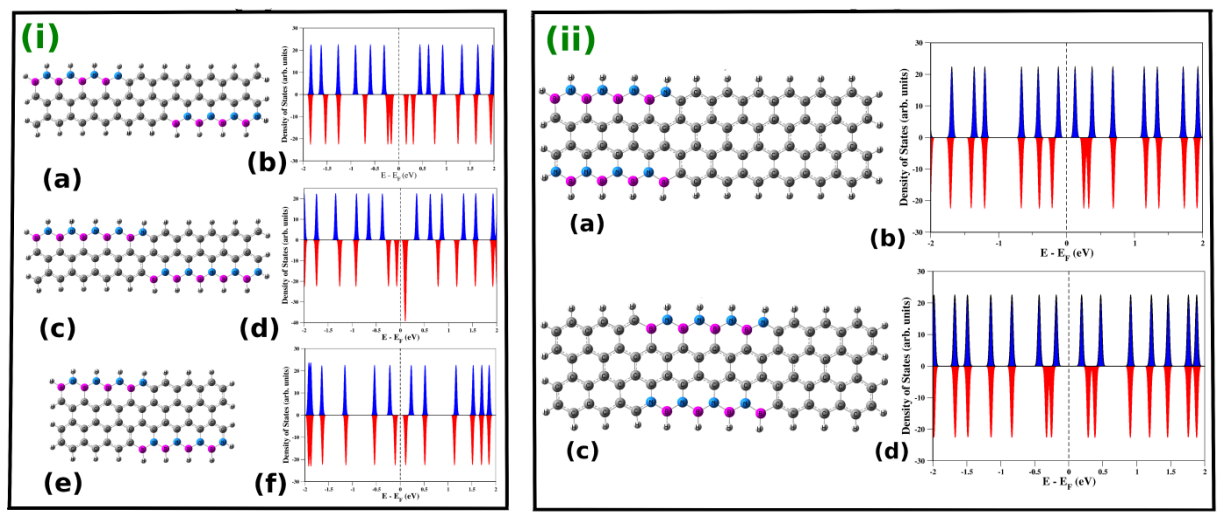

**Figure S4: (i)** Structures of the systems with (a) width less than that of, (c) substitution more than that of and (e) width and length less than that of (21, 4)-BN-partial-ed-GQD. (b), (d), (f) represents the corresponding DOS

plots manifesting the spin-polarized H-L gaps; **(ii)** (a), (c): systems with substituent atoms at symmetric positions of zigzag edges; (b), (d): DOS plots of (a), (c), respectively, emphasizing the spin-dependent H-L gap.

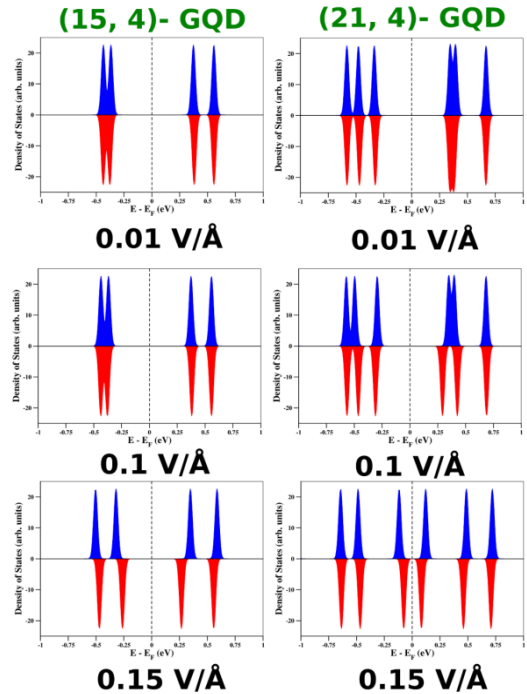

**Figure S5:** DOS plots of (15, 4) and (21, 4) GQDs at different electric-field strengths.

*Proving asymmetric substitution can't be the cause for spin-polarization*:

As we understood that the intrinsic electric-field is not the cause for the spin-dependent H-L gap, next, we looked at the other possible reasons. As mentioned earlier, H-L gap of (21, 4)-GQD is spin-dependent when it is partially substituted (as in figure 1b) and is spin-independent when it is completely substituted (as in figure 1c). One can interpret this result by observing these structures, at least, in two ways: (a) H-L gap is spin-dependent (spin-independent) when the position of the substituents at the zigzag edges is asymmetric (symmetric) and (b) H-L gap is spin-dependent when the system has both border and edge carbon atoms. To check the former interpretation we have considered two other structures as shown in figures S4(ii)a and S4(ii)c, in which the position of the substituents at the zigzag edges is symmetric and their DOS is shown in figures S4(ii)b and S4(ii)d, respectively. Clearly, these structures also show a spin-dependent H-L gap, ruling out the former interpretation.

*Proving the reason for spin-polarized H-L gap as "Presence of border and edge atoms":*

Now, if the latter interpretation is correct, then we have to find spin-dependent H-L gap for all the systems which have edge and border carbon atoms, irrespective of their number and their position. We have already proved that H-L gap is spin-dependent irrespective of the position of the edge and border carbon atoms (in figure 3(ii)). Next, to check whether H-L gap will continue to be spin-dependent with a change in the number of edge and border carbon atoms or not, we have gradually increased the amount of substitution at both the zigzag edges of (21, 4)-GQDs, and the results are given in the main article. All those reults strongly suggest that the reason for the spin-dependent H-L gap in BN-partial-ed-GQDs is due to the interplay between the edge and border carbon atoms.

**Table S1:** Mülliken population analysis of (21, 6) and (21, 8) BN-partial-ed-GQDs.

| (21, 6) BN-partial-ed-GQD | | | | | | | | (21, 8) BN-partial-ed-GQD | | | | | | | |
|---|---|---|---|---|---|---|---|---|---|---|---|---|---|---|---|
| B-edge | | N-edge | | B-border | | N-border | | B-edge | | N-edge | | B-border | | N-Border | |
| α-spin | β-spin | α-spin | β-spin | α-spin | β-spin | α-spin | β-spin | α-spin | β-spin | α-spin | β-spin | α-spin | β-spin | α-spin | β-spin |
| 10.83 | 11.97 | 11.99 | 10.86 | 7.28 | 7.73 | 8.3 | 7.89 | 12.04 | 10.77 | 10.8 | 10.06 | 7.16 | 7.84 | 7.42 | 7.78 |

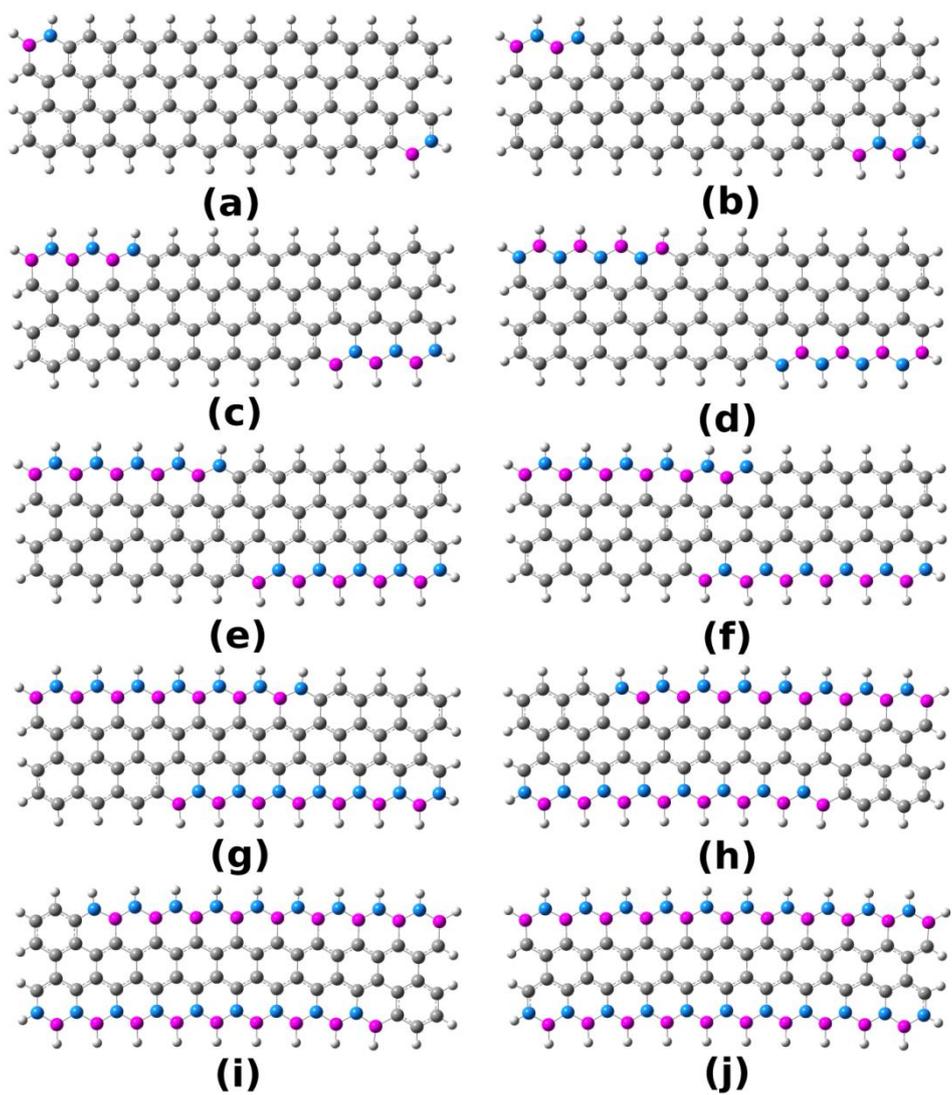

**Figure S6:** (a-j) shows structures with different amount of substitution

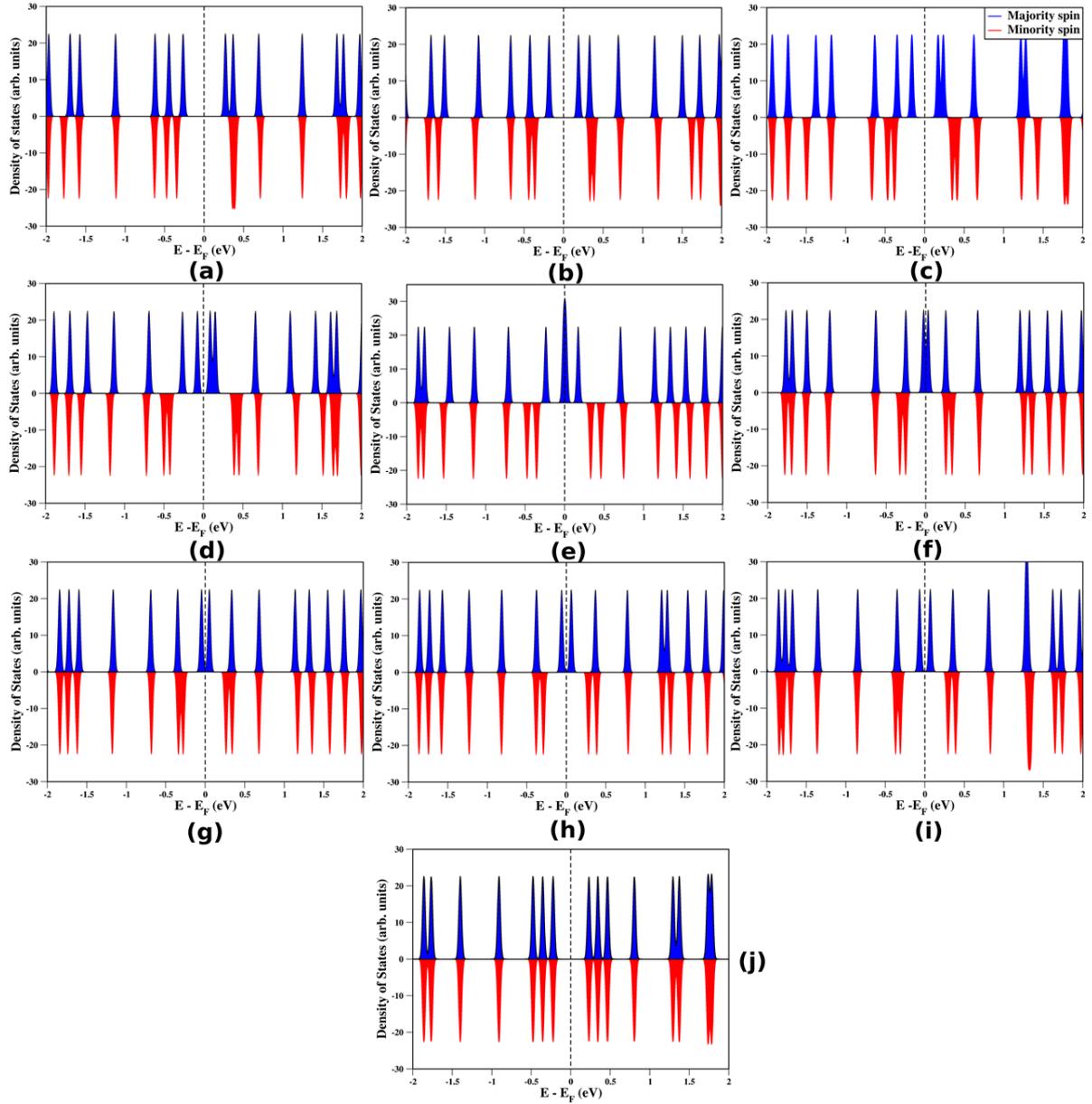

**Figure S7:** (a-j) shows DOS plots of the structures with different amount of substitution

The results of figure 3 (from the main article) can also be used to prove that there is no transition in the H-L gap from spin-dependency to spin-independency, with a change in the substitution across the zigzag edges (except for the complete substitution).

Also, the same results can be used to prove that intrinsic electric-field is not the cause for spin-dependent H-L gap in BN-half-ed-GQDs. This is because, when both the zigzag edges are completely substituted (the best way of substitution to generate intrinsic electric-field), we didn't observe any spin-dependent H-L gap (see figure S6-j), but, we could observe a spin-dependent H-L gap (see figure S6-i) when we remove just two substituent atoms (a structure which could also generate intrinsic electric-field, approximately in the same amount, as the previous one), at each edge from the previous system (see figure S5-i).

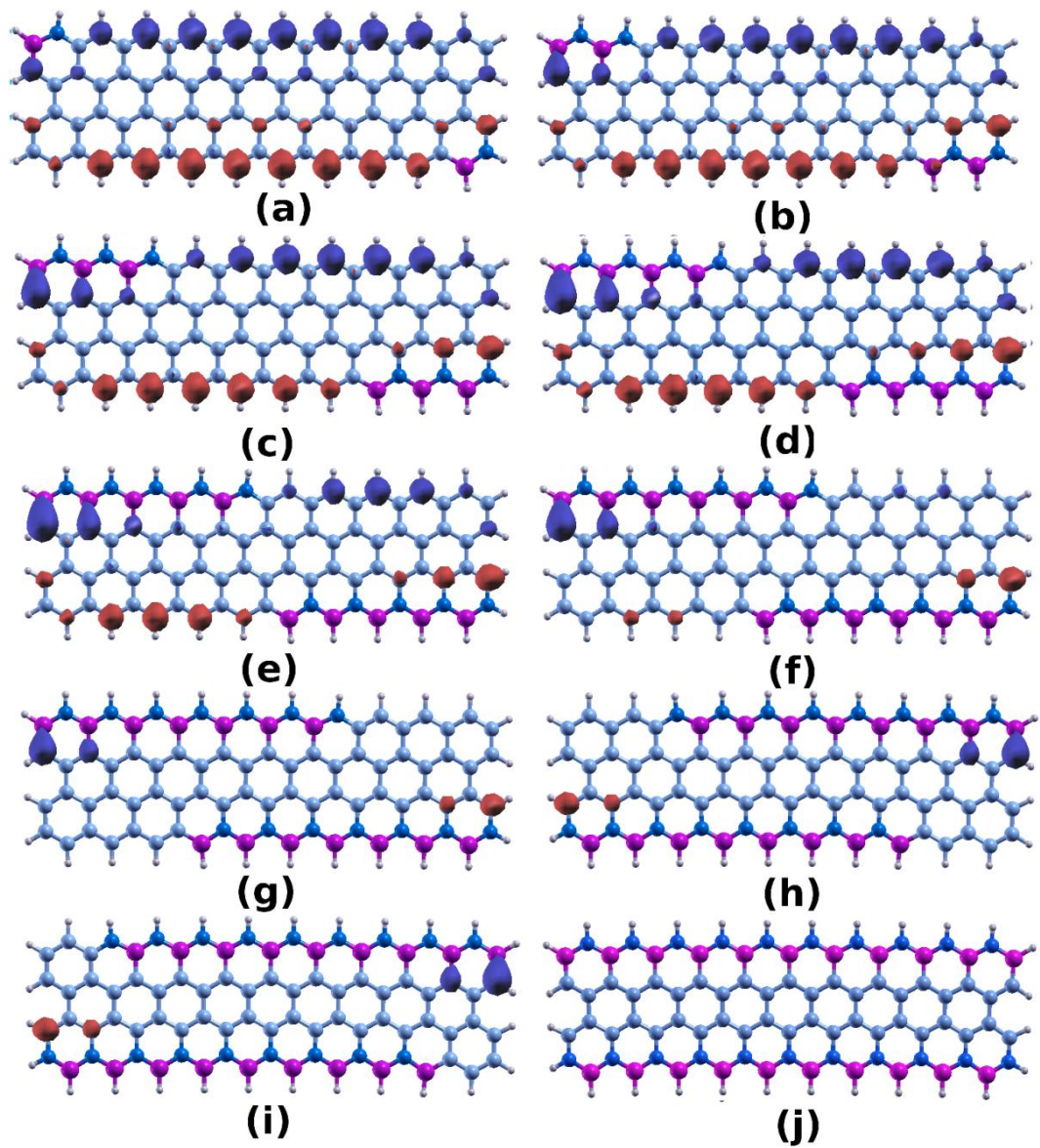

**Figure S8:** (a-j) shows spin density plots of all the structures with different amount of substitution

*Importance of boron atoms being at one edge and nitrogen atoms at the other for the observation of spin-polarized H-L gaps in BN-partial-ed-GQDs:*

The importance can be clearly understood by substituting both the edges with same atoms (i.e. either with boron or nitrogen). If we substitute both the edges with same atoms, then, despite of the spin-dependent charge-transfer at each border, there will not be any spin-dependency in the H-L gap for these systems, as shown in figure S9. This is because, the amount of the spin-polarization obtained at one border is exactly equal to the amount of spin-polarization obtained at the other border, but in opposite direction, leading to a zero total spin-polarization (see table S2) as observed in the case of GQDs. In fact, the spin-degeneracy in these systems is solely because of the spin-arrangement in the GQD (if spin-A at one edge, then spin-B at the other). Thus, not only the presence of border and edge carbon atoms, but the presence of different kinds of border and edge carbon atoms is the main reason for the observation of spin-dependent H-L gap in BN-parital-ed-GQDs.

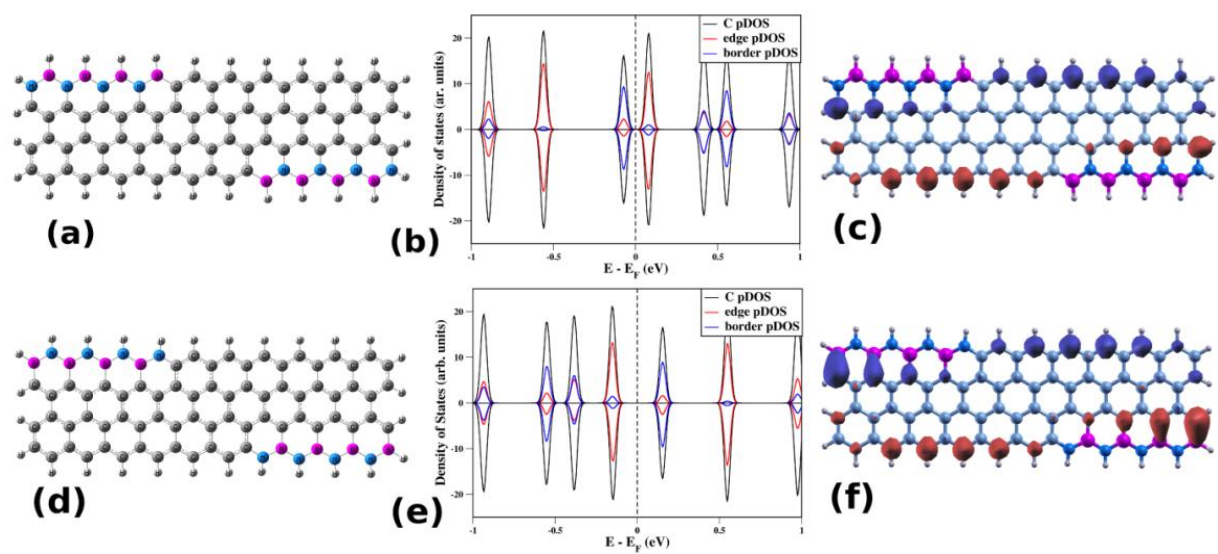

**Figure S9:** DOS, pDOS and spin distribution plots of GQDs with substituent atoms of same kind at each edge.

**Table S2:** Mülliken population analysis of (21, 4)- BN-parital-ed-GQDs with only one type of edge/border carbon atoms.

| BN-partial-ed-GQD (only B at edge) | | | | BN-partial-ed-GQD (only N at edge) | | | |
|---|---|---|---|---|---|---|---|
| 1st Border | | 2nd Border | | 1st Border | | 2nd Border | |
| α-spin | β-spin | α-spin | β-spin | α-spin | β-spin | α-spin | β-spin |
| 7.77 | 8.44 | 8.44 | 7.77 | 7.12 | 7.87 | 7.87 | 7.12 |

**Related to "D. Optical properties":**

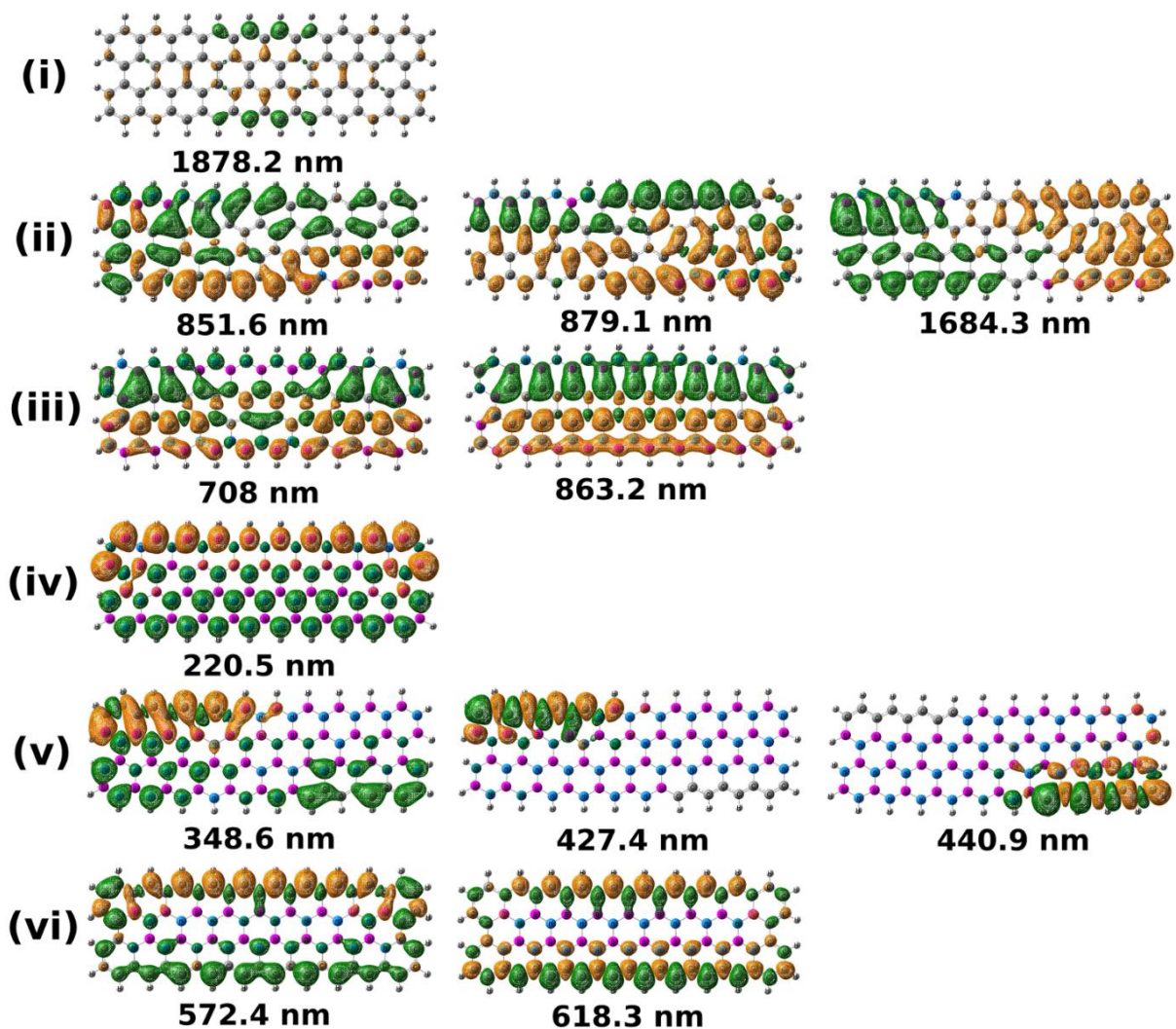

**Figure S10:** Electron density difference maps (EDDMs) of (21, 4)- QDs. The difference maps are calculated by subtracting the electron density on the QDs before and after the electronic transition. The figure corresponds to (i) GQD, (ii) BN-partial-ed-GQD (iii) BN-full-ed-GQD (iv) BNQD (v) C-partial-ed-BNQD and (vi) C-full-ed-BNQD. The corresponding wavelength values of the transitions are given bellow each figure. For all the plots, an isovalue of 0.0002 e/(Å)$^3$ is considered. The direction of electron-transfer is from the green region to orange region.

**Table S3:** The electronic transitions whose contributions are major in the absorption profile of the QDs.

| System | Wavelength (nm) | Osc. Strength | Major contributions |
|---|---|---|---|
| GQD | 1878.2 | 0.55 | H-1→LUMO (55%)<br>HOMO→L+1 (78%) |
| | 744.5 | 0.07 | H-2→L+2 (30%)<br>H-1→L+3 (42%)<br>HOMO→L+3 (19%) |
| BN-partial-ed-GQD | 1684.3 | 0.08 | H-1→L+1 (103%) |
| | 879.1 | 0.11 | HOMO→L+3 (86%)<br>H-1→L+2 (6%) |
| | 851.6 | 0.09 | H-3→LUMO (89%)<br>H-3→L+1 (2%) |
| BN-full-ed-GQD | 863.2 | 1.14 | H-1→LUMO (53%)<br>HOMO→L+1 (44%) |
| | 708 | 0.67 | H-2→L+1 (63%)<br>H-1→L+2 (32%) |
| | 556.3 | 0.04 | H-2→L+2 (30%)<br>H-1→L+3 (54%) |
| BNQD | 220.5 | 0.07 | H-3→L+3 (13%)<br>H-1→L+3 (26%)<br>HOMO→L+2 (41%) |
| C-partial-ed-BNQD | 427.4 | 0.81 | H-1→LUMO (95%)<br>H-3→LUMO (2%) |
| | 440.9 | 0.79 | HOMO→L+1 (91%)<br>HOMO→L+2 (3%) |
| | 348.6 | 0.39 | H-3→LUMO (60%)<br>H-2→LUMO (28%) |
| | 326.1 | 0.21 | H-2→L+1 (40%)<br>HOMO→L+7 (52%) |
| C-full-ed-BNQD | 618.3 | 2.78 | H-5→LUMO (10%)<br>H-2→LUMO (24%)<br>HOMO→L+3 (58%) |
| | 572.4 | 0.98 | H-5→LUMO (62%)<br>H-4→LUMO (16%) |
| | 571.8 | 0.27 | H-5→LUMO (16%)<br>H-4→LUMO (59%) |